\documentclass[useAMS,usenatbib]{mn2e}
\usepackage{epsfig}
\usepackage{amsmath}

\newcommand{\be}{\begin{equation}}
\newcommand{\beq}{\begin{equation}}
\newcommand{\ba}{\begin{eqnarray}}
\newcommand{\ee}{\end{equation}}
\newcommand{\eeq}{\end{equation}}
\newcommand{\ea}{\end{eqnarray}}

\newcommand{\hs}{\hspace{1mm}}

\newcommand{\apj}{ApJ}
\newcommand{\aap}{A\&A}
\newcommand{\apjl}{ApJL}
\newcommand{\mnras}{MNRAS}
\newcommand{\aj}{AJ}

\newcommand{\nat}{{\it Nature}}
\newcommand{\araa}{ARA\&A}
\newcommand{\pasj}{PASJ}

\def\lsim{~\rlap{$<$}{\lower 1.0ex\hbox{$\sim$}}}

\def\gsim{~\rlap{$>$}{\lower 1.0ex\hbox{$\sim$}}}

\title[Probing Reionization with the Ly$\alpha$ and UV Luminosity Functions]{Luminosity Functions of Ly$\alpha$ Emitting Galaxies and Cosmic Reionization of Hydrogen}

\author[Mark Dijkstra, Stuart Wyithe \& Zolt\'{a}n Haiman]{Mark Dijkstra$^{1}$\thanks{E-mail:dijkstra@physics.unimelb.edu.au}, J. Stuart B. Wyithe$^{1}$\thanks{E-mail:swyithe@physics.unimelb.edu.au} and Zolt\'{a}n Haiman$^{2}$\thanks{E-mail:zoltan@astro.columbia.edu}\\
$^{1}$School of Physics, University of Melbourne, Parkville, Victoria, 3010, Australia\\
$^{2}$ Astronomy Department, Columbia University, 550 West 120$^{\rm th}$ Street, New York, NY 10027, USA}

\def\LaTeX{L\kern-.36em\raise.3ex\hbox{a}\kern-.15em
    T\kern-.1667em\lower.7ex\hbox{E}\kern-.125emX}

\begin{document}

\date{\today}
\pagerange{\pageref{firstpage}--\pageref{lastpage}} \pubyear{2006}

\maketitle

\label{firstpage}

\begin{abstract}
Recent observations imply that the observed number counts of Ly$\alpha$ emitters (LAEs) evolved significantly between $z=5.7$ and $z=6.5$. It has been suggested that this was due to a rapid evolution in the ionisation state, and hence transmission of the IGM which caused Ly$\alpha$ flux from $z=6.5$ galaxies to be more strongly suppressed. In this paper we consider the joint evolution of the Ly$\alpha$ and UV luminosity functions (LFs) and show that the IGM transmission evolved between $z=6.5$ and $z=5.7$ by a factor $1.1 <R < 1.8$ ($ 95\%$ CL). This result is insensitive to the underlying model of the Ly$\alpha$ LF (as well as cosmic variance). Using a model for IGM transmission, we find that the evolution of the mean IGM density through cosmic expansion alone may result in a value for the ratio of transmissions as high as $R=1.3$. Thus, the existing LFs do not provide evidence for overlap. Furthermore, the constraint $R<1.8$ suggests that the Universe at $z=6.5$ was more than half ionised by volume, i.e. $x_{\rm i,V}>0.5$.
\end{abstract}

 
\section{Introduction}
\label{sec:intro}

Reionization of the Intergalactic Medium (IGM) is one of the milestones in the history of our Universe. Following recombination of the primordial plasma, gas in the Universe remained neutral until it became reionised by the first sources of UV-radiation. The {\it epoch of reionization} began with the formation of these first UV-sources, and ended when the entire IGM was cleared of neutral gas. The process of reionization was complex and is not well understood. The nature of the first UV-sources is unknown. Furthermore, various feedback mechanisms, both positive and negative, affect subsequent star, galaxy and black hole formation and hence the later stages of reionization. Extensive modelling, both semi-analytical \citep[e.g.][]{Cen03,WL03a,WL03b,HH03} and numerical \citep[e.g][]{Gnedin00,Iliev06}, has been performed, with the aim of interpreting the few existing observational constraints on reionization \citep[see][for a review]{Fan06rev}. The detection of flux blueward of the Ly$\alpha$ in high redshift quasars suggests that the Universe is fully ionised at $z \lsim 6$ \citep[e.g.][]{Fan02}. At higher redshifts, several Sloan Digital Sky Survey (SDSS) quasars show complete Gunn-Peterson troughs \citep{Fan06}. These troughs can be translated into lower limits on the neutral fraction of hydrogen in the IGM, and suggest that reionization may not have been completed until $z\approx6$ \citep{Fan02,Lidz02,WL04,MH04,MH06}. On the other hand, the measurement of optical depth for CMB photons to Thomson scattering by the {\it Wilkinson Microwave Anisotropy Probe} ({\it WMAP}) of $\tau_e=0.09 \pm 0.03$ \citep{wmap}, implies that the IGM is fully reionised out to $z=12 \pm 3$ or partially reionised out to even higher redshifts \citep[e.g. Fig~3 of][]{wmap}. 

Whether reionization indeed ended at $z\sim 6$ is still an open question. By adding the derived sizes of HII bubbles surrounding observed $z=6.5$ Ly$\alpha$ emitters, \citet{MR06} have shown that at least $20-50\%$ of the volume of the IGM had been reionised by $z=6.5$. \citet{Becker06} suggested that the observed Gunn-Peterson troughs in quasar spectra at $z\gsim 6$ can be explained without invoking any abrupt changes in the neutral fraction of the IGM \citep[also see][]{Ma06,BH07}. On the other hand, \citet{Ka06} have found the Ly$\alpha$ luminosity function (hereafter LF) to evolve between $z=5.7$ and $z=6.5$. In particular, the $z=6.5$ Ly$\alpha$ LF lies significantly below that measured at $z=5.7$. Furthermore, \citet{Ka06} show that the rest frame UV-LF of LAEs does not evolve, within its uncertainties, between $z=5.7$ and $z=6.5$. These observations are explained naturally, if the flux from $z=6.5$ Ly$\alpha$ emitters is attenuated by a larger factor than the flux from emitters at $z=5.7$ \citep{H02,S04}. The increase in attenuation could be interpreted as sudden change in the intergalactic neutral hydrogen content, which is thought to be a key feature of the end of the reionization epoch. Thus, the Ly$\alpha$ LF could be used to probe the last stages of reionization. In this paper we investigate the constraints that may be placed on the evolution of the IGM transmission using the observed Ly$\alpha$ LFs at $z=5.7$ and $z=6.5$. As part of our analysis, we also investigate how the rest-frame UV-LF of LAEs may provide additional constraints.

\citet{HS99} were the first to advocate the use of the Ly$\alpha$ luminosity function to constrain reionization, and several other papers have followed since \citep[e.g.][]{MR04,D05,HC05,D06,F06,Mao06}. The work presented in this paper differs from other investigations in two important aspects. Firstly, our model incorporates detailed calculations of the IGM transmission \citep{paperI}. For a description of these calculations the reader is referred to that paper. Here, we briefly summarise the models main ingredients. The IGM transmission is calculated using a model for the gas in the IGM that accounts for clumping, and infall. In this model, resonant absorption of Ly$\alpha$ photons by gas in the infall region (which extends out to several virial radii) erases a significant fraction of the Ly$\alpha$ line. The model also accounts for damping wing absorption in cases in which a (partially) neutral IGM surrounds the HII bubble. Secondly, we perform statistical comparisons to the most up-to date observations.

In \S~\ref{sec:model} we present our model of the Ly$\alpha$ LF. In \S~\ref{sec:data} we describe the data available. The results of our comparison with the data and conclusions are presented in \S~\ref{sec:result} and \S~\ref{sec:conclusion}, respectively. The parameters for the background cosmology used throughout this paper are $\Omega_m=0.24$, $\Omega_{\Lambda}=0.76$, $\Omega_b=0.044$, $h=0.73$ and $\sigma_8=0.74$ \citep{wmap}. 

\section{The Model of the Ly$\alpha$ Luminosity Function}
\label{sec:model}
\begin{figure*}
\vbox{\centerline{\epsfig{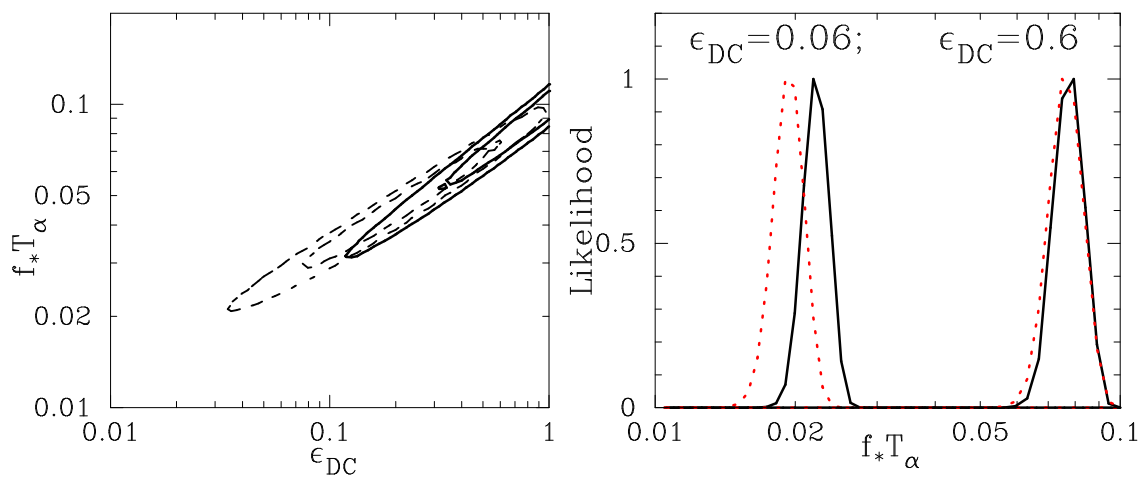}}}
\caption[]{{\it Left Panel:} Joint constraints on $\epsilon_{\rm DC}$ and $f_*\mathcal{T}_{\alpha}$. The likelihood contours at $64\%$, $26\%$ and $10\%$ of the peak likelihood are shown at $z=5.7$ as {\it dashed} contours with those at $z=6.5$ overlaid as {\it solid} contours. Although the likelihood peaks at different values for $\epsilon_{\rm DC}$ and $f_*\mathcal{T}_{\alpha}$ at $z=5.7$ and $z=6.5$, the contours at both redshifts fill the same region of the $(\epsilon_{\rm DC},f_*\mathcal{T}_{\alpha})$-plane. A strong degeneracy exists between $\epsilon_{\rm DC}$ and $f_*\mathcal{T}_{\alpha}$ (see text). {\it Right Panel:} Slices through the likelihood surface at $\epsilon_{\rm DC}=0.06$ and $\epsilon_{\rm DC}=0.06$ are shown at $z=5.7$ ({\it black solid lines}) and at $z=6.5$ ({\it red dotted lines}), normalised to a peak of unity. For a given $\epsilon_{\rm DC}$, $f_*\mathcal{T}_{\alpha}$ peaks at very similar values at both redshifts. Assuming that $f_*$ remains constant between $z=5.7$ and $z=6.5$, this suggests that the IGM transmission evolves only weakly between $z=5.7$ and $z=6.5$.}
\label{fig:p}
\end{figure*}
We use the following simple prescription to relate the Ly$\alpha$ luminosity of a galaxy to the mass of its host dark matter $M_{\rm tot}$. The total mass of baryons within a galaxy is $(\Omega_b/\Omega_m)M_{\rm tot}$, of which a fraction $f_*$ is assumed to be converted into stars over a time scale of $t_{\rm sys}=\epsilon_{\rm DC}t_{\rm hub}$. Here, $\epsilon_{\rm DC}$ is the duty cycle and $t_{\rm hub}(z)\equiv \frac{2}{3H(z)}$, is the Hubble time at redshift $z$. This prescription yields a star formation rate of $\dot{M}_*=f_*(\Omega_b/\Omega_m)M/t_{\rm sys}$. The star formation rate can be converted into a Ly$\alpha$ luminosity by assuming that approximately two-thirds of the ionising photons absorbed within the galaxy are converted into Ly$\alpha$ \citep{osterbrock}, thus $L_{\alpha}=0.68h\nu_{\alpha}(1-f_{\rm esc})\dot{Q}_{\rm H}$. Here, $f_{\rm esc}$ is the escape fraction of ionising photons, $f_{\rm esc}\ll 1$, and $h\nu_{\alpha}=10.2$ eV is the energy of a Ly$\alpha$ photon. The total output of ionising photons per unit mass of star formation, $\dot{Q}_{\rm H}$, depends on the metallicity of the gas from which stars formed, as well as their initial mass function \citep[][]{K98,S03}. In the reminder of this paper, we assume a Salpeter IMF and $Z=0.05Z_{\odot}$.\footnote{The uncertainty in the relation between Ly$\alpha$ luminosity and $\dot{M}_*$ is large, as it depends on an unknown IMF and metallicity of the gas from which the galaxies formed \citep{S03}. For the present work, because of the degeneracy between $f_*$ and the prefactor in Eq~2., the exact relation (and therefore the assumed metallicity) is not important.} The number density of Ly$\alpha$ emitters with Ly$\alpha$ luminosities exceeding $\mathcal{T}_{\alpha}\times L_{\alpha}$ is then given by
\begin{equation}
N(>\mathcal{T}_{\alpha}\times L_{\alpha})=\epsilon_{\rm DC}\int_{M_{\alpha}}^{\infty}dM\frac{dn}{dM},
\label{eq:form} 
\end{equation}where the Ly$\alpha$ luminosity and host halo mass, $M_{\alpha}$ are related\footnote{Although it is reasonable to assume that the star formation rate-and thus Ly$\alpha$ luminosity-in a galaxy is related to the total amount of gas available, and thus to $M_{\rm tot}$, no direct observational evidence exists to support this assumption. The fact that we obtain good fits to the data for reasonable model parameters (\S~\ref{sec:result}) is promising, and suggests that our model provides  a reasonable description. However, even if one dismisses the assumption that Ly$\alpha$ luminosity relates to halo mass, then it is still possible to obtain constraints on the change of the IGM transmission with redshift using the rest-frame UV-LF of LAEs (as is discussed in \S~\ref{sec:UV}).} by
\begin{equation}
\mathcal{T}_{\alpha} \times L_{\alpha}=2.0 \times 10^{42}\hs{\rm erg} \hs{\rm s}^{-1}\frac{M_{\alpha}(M_{\odot}) \frac{\Omega_b}{\Omega_m}f_*}{t_{\rm sys}({\rm yr})}\mathcal{T}_{\alpha}.
\label{eq:2}
 \end{equation} In this relation, $\mathcal{T}_{\alpha}$ is the IGM transmission multiplied by the escape fraction of Ly$\alpha$ photons from the galaxy. The function $dn/dM$ is the Press-Schechter (1974) mass function \citep[with the modification of][]{ST01}, which gives the number density of halos of mass $M$ (in units of comoving Mpc$^{-3}$).

According to this prescription, the Ly$\alpha$ LF has two free parameters $(\epsilon_{\rm DC},f_*\mathcal{T}_{\alpha})$. Thus, we explicitly assume that $T_{\alpha}$ is independent of $M_{\alpha}$ and $\dot{M}_*$. In \citet{paperI}, we found $T_{\alpha}$ to decrease with increasing $M_{\alpha}$ (because infall is more prominent around higher mass halos) and increasing $\dot{M}_*$ (because a larger output of ionising photons reduces the impact of resonant absorption in the infall region). Conversely, for a fixed $\epsilon_{\rm DC}$, increasing $M_{\alpha}$ results in an increase of $\dot{M}_*$, and hence an increase in $\mathcal{T}_{\alpha}$.  Given that the range of observed luminosities, and hence halo masses, spans only $\sim 1$ dex (\S~\ref{sec:data}), the variation in $\mathcal{T}_{\alpha}$ with $M_{\alpha}$ and $\dot{M}_*$ can be ignored. In this paper we compare models for the Ly$\alpha$ LF to observation as a function of these parameters and constrain the evolution in $f_*\mathcal{T}_{\alpha}$ between $z=5.7$ and $z=6.5$.  

\section{The Data}
\label{sec:data}
Detailed Ly$\alpha$ LFs have recently been presented by Shimasaku et al. (2006) for LAEs observed at $z=5.7$, and by Taniguchi et al. (2005) and Kashikawa et al. (2006), for LAEs observed at $z=6.5$. Some caution must be exercised when comparing our models to this data. First, the observed luminosities have been derived from the observed fluxes by assuming that all Ly$\alpha$ emerging from the galaxy was transmitted by the IGM. However, this is very unlikely. In fact, it has been shown that under reasonable model assumptions, the IGM transmits only a fraction of $\mathcal{T}_{\alpha}=0.1-0.3$ \citep{paperI} of Ly$\alpha$ photons, even when highly ionised. We therefore replace the quoted observed luminosities, $L_{{\rm Ly}\alpha}$ by $\mathcal{T}_{\alpha}\times L_{{\rm Ly}\alpha}$(as in Eq~\ref{eq:form} and Eq~\ref{eq:2}). Second, the error bars on the data are uncertain. For example, the {\it open circles} shown in our Figure~\ref{fig:lumfunc} and Figure~5 of \citet{Ka06} denote the LF derived from the raw counts of their spectroscopic sample+additional photometric sample. The {\it filled circles} are corrected for detection incompleteness. The error-bars on the {\it filled circles} denote Poisson-errors. Additionally, cosmic variance is expected to add a variance of $30\% (50\%)$ at the lowest (highest) luminosities \citep[e.g.][]{So04} and enters as an overall shift of the Ly$\alpha$ LF in the vertical direction. As a result, cosmic variance could be (partly) responsible for the off-set in the LFs between $z=5.7$ and $z=6.5$. In the first part of our analysis we consider the Ly$\alpha$ LF alone, and ignore cosmic variance. Allowing for cosmic variance would enlarge the errobars on model parameters and strengthen the conclusion that the evolution of the Ly$\alpha$ LF alone, can be fully accounted for by the evolution of the mass function of dark matter halos. In the second part, we remove the effect of cosmic variance by considering the joint evolution of the Ly$\alpha$ and UV-LFs.

Furthermore, the lower luminosity points ($\mathcal{T}_{\alpha}L_{{\rm Ly}\alpha}\leq 5 \times 10^{42}$ erg s$^{-1}$) suffer more from detection incompleteness, as evidenced by the larger offset between the corrected and uncorrected data points. In addition, the IGM is expected to transmit only 10-40\% of the Ly$\alpha$ flux for these galaxies, and the detection of these low luminosity Ly$\alpha$ emitters requires an unusually large IGM transmission \citep[$\mathcal{T}_{\alpha}\sim 1$, or unusually large Ly$\alpha$ luminosity][]{paperI}. It follows then that these galaxies must comprise only a (small) subset of the true sample. However, the exact uncertainty this introduces is not known. We account for this uncertainty by enlarging the adopted error-bars at $\mathcal{T}_{\alpha}L_{{\rm Ly}\alpha}\leq 5 \times 10^{42}$ erg s$^{-1}$ by factors of $4$ and $3$ for the lowest and second lowest luminosity point, respectively (on the {\it filled circles}). These enlargements were chosen so that the $1-\sigma$ error bars enclose the data points prior to the adjustment. The choice of error bars on these low luminosity points affects the values of our best fit model parameters, but not our conclusions regarding evolution in transmission of the IGM.

\section{Results}
\label{sec:result}

We calculated the Ly$\alpha$ LF for a grid of models in the $(\epsilon_{\rm DC},f_*\mathcal{T}_{\alpha})$-plane, and generated likelihoods $\mathcal{L}[P]={\rm exp}[-0.5\chi^2]$, where $\chi^2=\sum_i^{N_{\rm data}}({\rm model}_i-{\rm data}_i)^2/\sigma^2_{i}$, for each model. Here, data$_i$ and $\sigma_i$ are the i$^{\rm th}$ data point and its error, and model$_i$ is the model evaluated at the $i^{\rm th}$ luminosity bin. The sum is over $N_{\rm data}=6$ points.\footnote{The data points represent a cumulative number of galaxies with Ly$\alpha$ luminosities exceeding $L_{{\rm Ly}\alpha}$. The data points are therefore not completely independent. However, the number of galaxies more than doubles between most bins, and the errorbars do not overlap. This indicates that although cumulative, the points are close to independent.} 

Likelihoods were determined for each redshift independently. In the {\it left panel} of Figure~\ref{fig:p} we show likelihood contours in the $(\epsilon_{\rm DC},f_*\mathcal{T}_{\alpha})$-plane at $64\%$, $26\%$ and $10\%$ of the peak likelihood at $z=5.7$ (as {\it dashed} contours) and $z=6.5$ (overlaid as {\it solid} contours). The likelihood contours at both redshifts fill the same region of the $(\epsilon_{\rm DC},f_*\mathcal{T}_{\alpha})$-plane. The likelihoods peak at $(\epsilon_{\rm DC},f_*\mathcal{T}_{\alpha})\sim(0.2,0.03)$ at $z=5.7$ and at $(\epsilon_{\rm DC},f_*\mathcal{T}_{\alpha})=(1.0,0.1)$ at $z=6.5$, with a strong degeneracy between $\epsilon_{\rm DC}$ and $f_*\mathcal{T}_{\alpha}$. This degeneracy arises because a lower duty cycle corresponds to a lower $t_{\rm sys}$, which requires a lower $f_{*}$ to achieve the same star formation rate for a fixed value of $M_{\rm tot}$.
\footnote{Reducing $\epsilon_{\rm DC}$ and $f_*\mathcal{T}_{\alpha}$ by the same factor yields the same Ly$\alpha$ luminosity. However, because the galaxies are visible for a shorter time, this still results in the prediction of fewer objects. This causes the deviation from a 45-degree degeneracy-line in the $(\epsilon_{\rm DC},f_*\mathcal{T}_{\alpha})$-plane: changing $\epsilon_{\rm DC}$ by 1 order of magnitude is compensated for by changing $f_*\mathcal{T}_{\alpha}$ only by half an order of magnitude.}
\begin{figure}
\vbox{\centerline{\epsfig{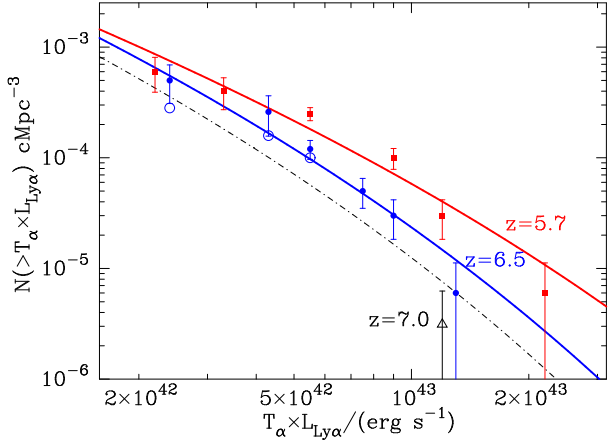}}}
\caption[]{The Ly$\alpha$ luminosity function (LF) at $z=5.7$ and $z=6.5$. The {\it red squares} and {\it blue circles} represent data from Shimasaku et al (2006, $z=5.7$) and Kashikawa et al (2006, $z=6.5$). Overplotted as {\it solid lines} are the best fit models. For completeness, we have shown the $z=7.0$ LF ({\it black dashed line}), taking our best fit model to the $z=6.5$ data and evolving only the mass function between $z=7.0$ and $z=6.5$. The {\it triangle} represents the $z=7.0$ galaxy discovered by \citet{Iye06}, which is consistent with the model LF.}
\label{fig:lumfunc}
\end{figure}

In Figure~\ref{fig:lumfunc} we show the best fit models at $z=6.5$ and $z=5.7$. Clearly, the models provide an adequate fit to the data. We also show the model $z=7.0$ LF ({\it black dot-dashed line}), under the assumption that only the halo mass function evolved between $z=6.5$ and $z=7.0$. The {\it triangle} represents the density implied by the $z=7.0$ galaxy discovered by \citet{Iye06}. This density is consistent with the model LF at the $1-\sigma$ level, suggesting that no drastic change in the IGM transmission is required between $z=6.5$ and $z=7.0$.

At $z=5.7$ the duty cycle was found to lie in the range $\epsilon_{\rm DC}=[0.03,1.0]$, while at $z=6.5$ this range was $\epsilon_{\rm DC}=[0.1,1]$. The likelihood in $\epsilon_{\rm DC}$ therefore extends over more than 1 dex in each case. As a result, the a-posteriori probability for $\epsilon_{\rm DC}$ is quite sensitive to the choice of a prior probability. Thus the data do not really constrain the duty cycle. Therefore, in the {\it right panel} of Figure~\ref{fig:p}, we show slices through the likelihood surface at $\epsilon_{\rm DC}=0.06$ and at $\epsilon_{\rm DC}=0.6$ respectively, normalised to a peak of unity. The distribution of $f_*\mathcal{T}_{\alpha}$ at $z=5.7$ and $z=6.5$ are shown as the {\it black solid} and {\it red dotted lines}. For fixed $\epsilon_{\rm DC}$, the most likely values of $f_*\mathcal{T}_{\alpha}$ are common at both redshifts. Assuming that $f_*$ does not evolve between $z=5.7$ and $z=6.5$, it therefore follows that the IGM transmission need not evolve significantly between $z=5.7$ and $z=6.5$ in order to explain the evolution of the LF. Thus the evolution in the observed LF may be attributed to the mass function of dark matter halos alone. This is explored in more detail in \S~\ref{sec:evo}.
 
Note that the best fit values for $f_*\mathcal{T}_{\alpha}=[0.01-0.05]$ meet our prior expectations. For $f_*=0.1$, we find $\mathcal{T}_{\alpha}=0.1-0.5$, which corresponds to the range of transmissions found in \citet{paperI}. In our best-fit models, the mass range of Ly$\alpha$ emitters is $M_{\rm tot}\sim 4-42 \times 10^{10}M_{\odot}$ at $z=5.7$ and $M_{\rm tot}\sim 6-32 \times 10^{10}M_{\odot}$ at $z=6.5$.
 
\subsection{Evolution of IGM Transmission}
\label{sec:evo}

In this section we investigate the evolution of the IGM transmission, $\mathcal{T}_{\alpha}$ in more detail. We found that the observed LFs can be described by models over a large range in duty cycle $\epsilon_{\rm DC}$. Here, we fix the duty cycle and $f_*$ to be common between $z=5.7$ and $z=6.5$, and vary the parameters $f_*\mathcal{T}_{\alpha,57}$ and $R\equiv \mathcal{T}_{\alpha,57}/\mathcal{T}_{\alpha,65}$, where $\mathcal{T}_{\alpha,57}$ and $\mathcal{T}_{\alpha,65}$ are the IGM transmission at $z=5.7$ and $z=6.5$, respectively. We compute model LFs on a grid in the $(R,f_*\mathcal{T}_{\alpha,57})$-plane and simultaneously fit to the observed LFs at $z=5.7$ and $z=6.5$. 

\begin{figure}
\vbox{\centerline{\epsfig{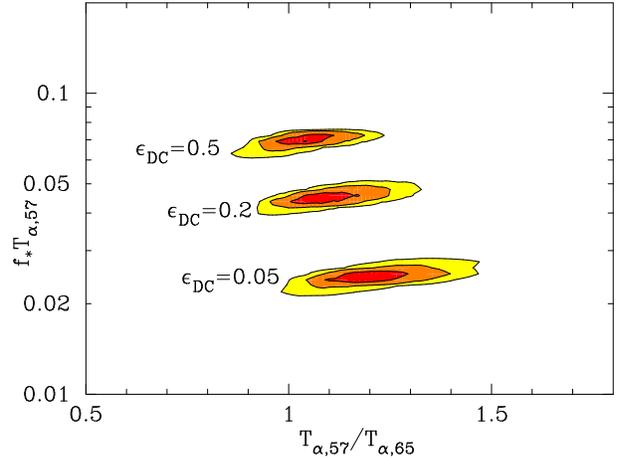}}}
\caption[]{Likelihood contours at $64\%$, $26\%$ and $10\%$ of the peak likelihood in the $(\mathcal{T}_{\alpha,57}/\mathcal{T}_{\alpha,65},f_*\mathcal{T}_{\alpha,57})$-plane, for three different duty cycles $\epsilon_{\rm DC}$. The models that provide the best fit to the observed LFs have $\mathcal{T}_{\alpha,57}/\mathcal{T}_{\alpha,65}\sim 0.8-1.5$. Therefore, the best fit-models do not favour a large change in the opacity of the IGM to Ly$\alpha$ photons emitted by galaxies.}
\label{fig:rattrans}
\end{figure}

The results of this calculation are shown in Figure~\ref{fig:rattrans}. Here, likelihood contours at $64\%$, $26\%$ and $10\%$ of the peak likelihood are plotted in the $(R,f_*\mathcal{T}_{\alpha,57})$-plane for 3 different duty cycles ($\epsilon_{\rm DC}=0.01,0.1$ and $0.5$). The likelihood-curves shift towards higher $f_*\mathcal{T}_{\alpha,57}$ for larger values of $\epsilon_{\rm DC}$ due to the degeneracy noted above.
This figure shows that the ratio $R\equiv \mathcal{T}_{\alpha,57}/\mathcal{T}_{\alpha,65}$ is close to $1$ (within $\lsim 2-\sigma$) for all reasonable values of $\epsilon_{\rm DC}$. Therefore, the observed evolution in the Ly$\alpha$ LF can be attributed to the evolution of the mass function of dark matter halos.

\subsection{Constraints on Transmission from the Ly$\alpha$ and UV-Luminosity Functions.}
\label{sec:UV}

In the previous section we derived constraints on the ratio $\mathcal{T}_{\alpha,65}/\mathcal{T}_{\alpha,57}$ using a simple model for the Ly$\alpha$ LF in which the star formation rate in a galaxy increases in proportion to $M_{\rm tot}$, the total mass of its host dark matter halo. We showed that the observed evolution in the Ly$\alpha$ LF can be attributed to the evolution of the mass function of dark matter halos. As evidenced by Figure~\ref{fig:rattrans}, our exact constraints are somewhat dependent on the unknown  model parameter $\epsilon_{\rm DC}$. The goal of this section is to obtain constraints on $\mathcal{T}_{\alpha}$ that are independent of any of our model parameters, and more generally, to obtain constraints that are independent of the model that underlies the Ly$\alpha$ LF.

\begin{figure}
\vbox{\centerline{\epsfig{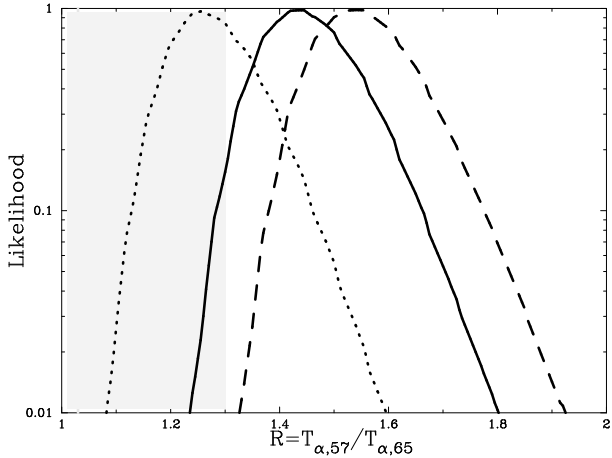}}}
\caption[]{The likelihood distribution of the ratio $\mathcal{T}_{\alpha,65}/\mathcal{T}_{\alpha,57}$ derived from the Ly$\alpha$ + rest-frame UV luminosity function. Three curves are shown: the {\it solid line} shows the likelihood under the assumption that the UV-LFs of LAEs are identical at $z=6.5$ and $z=5.7$. The {\it dotted/dashed} line shows the likelihood under the assumption that the UV-LFs of LAEs at $z=6.5$ lies higher by a factor of $0.7$ and $1.2$, respectively. The {\it grey thick region} at $R<1.3$ shows the range of ratios that can be expected if the Universe remained fully reionised between $z=6.5$ and $z=5.7$ (\S~\ref{sec:ratio}).}
\label{fig:uv}
\end{figure}

\citet{Ka06} found that the rest-frame UV-LF of LAEs does not evolve between $z=5.7$ and $z=6.5$. This appears to contradict the conclusion that the evolution of dark matter halos (possibly in combination with cosmic variance) caused the observed evolution of the Ly$\alpha$ LF. Indeed, if the rest-frame UV-LF of LAEs is identical at $z=5.7$ and $z=6.5$, then the entire difference between the Ly$\alpha$ LFs at $z=5.7$ and $z=6.5$ must be due to a change in either $\mathcal{T}_{\alpha}$ or in the ratio $L_\alpha/L_{\rm UV}$ (where $L_{\rm UV}$ is rest-frame UV luminosity of a galaxy). In other words, a lower IGM transmission (or $L_\alpha/L_{\rm UV}$ ratio) at $z=6.5$ would shift the Ly$\alpha$ luminosity function to the left relative to the $z=5.7$ Ly$\alpha$ LF. Assuming for the moment that $L_\alpha/L_{\rm UV}$ remains constant, we quantify the ratio $R\equiv \mathcal{T}_{\alpha,65}/\mathcal{T}_{\alpha,57}$ that is favored by this constraint. While the analysis could be done for individual galaxies, we consider the sample as a whole, by taking the model parameters $(f_*,\epsilon_{\rm DC})$ of our best-fit $z=5.7$ model. We then obtain $z=6.5$ Ly$\alpha$ LFs for a range of $\mathcal{T}_{\alpha,65}$ by scaling the luminosities in the $z=5.7$ model by a factor of $\mathcal{T}_{\alpha,65}/\mathcal{T}_{\alpha,57}$. In Figure~\ref{fig:uv} we plot the likelihood, $\mathcal{L}[P]={\rm exp}[-0.5\chi^2]$, as function of the ratio $\mathcal{T}_{\alpha,65}/\mathcal{T}_{\alpha,57}$.
 If the UV-LF remained constant between $z=5.7$ and $z=6.5$, then we find $R=1.4^{+0.3}_{-0.1}$ ($95\%$ CL). The preferred ratio decreases if the UV-LF at $z=6.5$ lies below that at $z=5.7$. Existing data is inconclusive regarding the precise evolution of the UV-LF between $z=5.7$ and $z=6.5$: from the size of the error-bars in Figure~7 of \citet{Ka06}, we find that the best-fit $z=6.5$ UV-LF to lie higher by a factor of $0.7\pm0.2$ (95\%) than at $z=5.7$ if all data points brighter than $M_{\rm UV}=-20$ are used, while it lies higher by a factor of $1.2 \pm 0.4$ (95\%) if the two data points at $M_{\rm UV}\sim -20.2$ are ignored. Figure~\ref{fig:uv} shows the the likelihood of the ratio $\mathcal{T}_{\alpha,65}/\mathcal{T}_{\alpha,57}$ assuming that the $z=6.5$ LF lies higher by a factor of $0.7$ (1.2) as the {\it dotted (dashed) line}. Assuming for simplicity that the ratio of the $z=6.5$ and $z=5.7$ UV-LFs has a flat likelihood distribution between $0.7$ and $1.2$, then it follows from Figure~\ref{fig:uv} that $1.1 < R < 1.8$ ($\sim 95\%$ CL). We quantify in more detail below how this compares to the evolution of the transmission in a fully (\S~\ref{sec:ratio}) and partially (\S~\ref{sec:bub}) reionised universe.

The relative normalisations of the UV \& Ly$\alpha$ LFs may evolve for reasons other than transmission. Either (1) $L_{\rm UV}$ increases with redshift, or (2) $L_\alpha$ decreases with redshift. (1) could be caused by decreasing dust abundance towards higher redshift. However, the observed Ly$\alpha$ LF requires that Ly$\alpha$ is affected less by this possible evolution of the dust content, which in turn requires fine tuning of how the dust is distributed. (2) could be caused by an increasing escape fraction of ionising photons with redshift as $L_{\alpha} \propto (1-f_{\rm esc})$. Recently, \citet{Inoue06} found that $f_{\rm esc}$ is roughly consistent with the value $f_{\rm esc}\sim 0.1$ at $4 \lsim z \lsim 6$. In order for $L_{\alpha}$ to in increase by a factor of $\geq 1.1$ between $z=6.5$ and $z=5.7$ would require $f_{\rm esc}\geq 0.2$ at $z=6.5$.

\subsection{Cosmic Expansion and the Evolution of $\mathcal{T}_{\alpha}$}
\label{sec:ratio}
\begin{figure*}
\vbox{\centerline{\epsfig{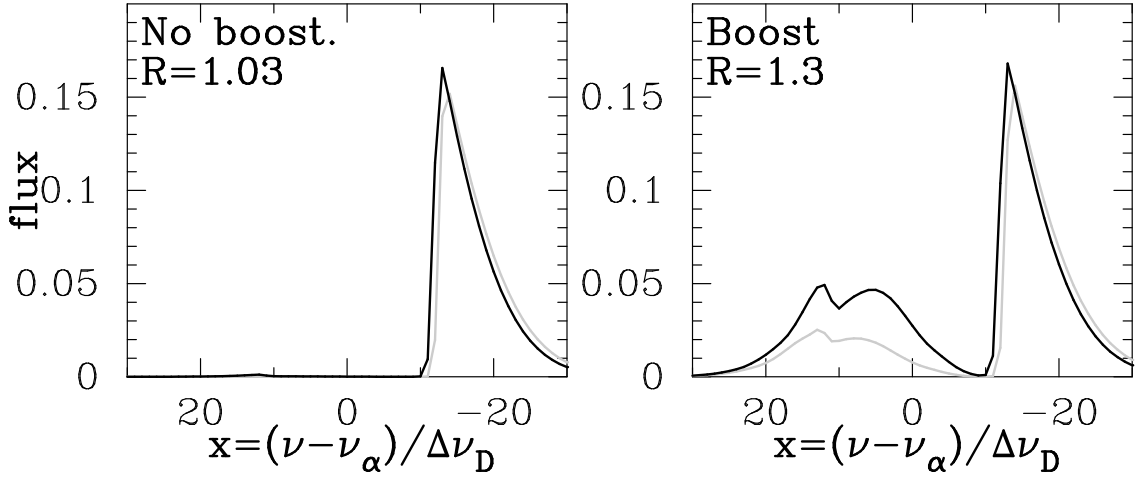}}}
\caption[]{Ly$\alpha$ spectra of LAEs at $z=5.7$ ({\it black line}) and $z=5.7$ ({\it grey line}) after processing through a fully reionised IGM. In the {\it left/right panel} the local ionising background is not/is enhanced due to the clustering of nearby undetected sources. The figure shows that $R\equiv \mathcal{T}_{\alpha,57}/\mathcal{T}_{\alpha,65}$ may be as large as $R=1.3$ without invoking any drastic change in the ionisation state of the IGM.}
\label{fig:ratio}
\end{figure*}
Following the analysis presented in \citet{paperI}, we calculate the IGM transmission ($\mathcal{T}_{\alpha}$) at $z=5.7$ and $z=6.5$, and the expected value of the ratio of transmission $R\equiv \mathcal{T}_{\alpha,57}/\mathcal{T}_{\alpha,65}$. We calculate $\mathcal{T}_{\alpha}$ for a model in which $M_{\rm tot}=10^{11}M_{\odot}$ (the centroid of our best-fit mass range for the host halos of the LAEs \S~\ref{sec:result}), and $\dot{M}_*=10M_{\odot}$/yr. The Ly$\alpha$ line before scattering in the IGM is assumed to be Gaussian with a standard deviation of $v_{\rm circ}$. The density and velocity profiles of gas in the IGM are described in \citet{paperI}. Note that the precise value of $\mathcal{T}_{\alpha}$ differs with quantities such as $M_{\rm tot}$, $\dot{M}_*$, and other parameters such as the assumed width of the Ly$\alpha$ line, and the peculiar velocity of the galaxies relative to the surrounding IGM \citep{paperI}. However, as long as these quantities do not vary significantly between the samples at $z=5.7$ and $z=6.5$, these uncertainties should not strongly affect the calculation of the ratio $R$. 

We consider two cases: in case I the photoionisation rate due to the externally generated, ionising background is set to $10^{-13}$ s$^{-1}$ as derived from quasar absorption spectra \citep[e.g.][]{Fan06}. In case II clustering of nearby, undetected, sources boosts the photoionisation rate considerably. We include the effects of clustering using the prescription in \citet{paperI}. Clustering of sources and the corresponding boost in the local ionising background occurs naturally in hierchical models \citep{WL05}. In both cases, the only difference between models at $z=5.7$ and $z=6.5$ is the mean density of baryons in the universe. Figure~\ref{fig:ratio} shows the observed Ly$\alpha$ lines after processing through the IGM. The horizontal axis shows the normalized frequency\footnote{Here $x\equiv (\nu-\nu_\alpha)/\Delta \nu_D$, where $\Delta \nu_D\equiv \nu_\alpha v_{\rm th}/c$. Here $v_{th}=\sqrt{2k_BT/m_p}$ is the thermal velocity of the hydrogen atoms in the gas, $k_B$ is the Boltzmann constant, $T$ the gas temperature, $m_p$ the proton mass.} $x\equiv (\nu-\nu_\alpha)/\Delta \nu_D$. Note that $x=0$ corresponds to the true line center. The flux density on the horizontal axis is in arbitrary units. Here, case I (no boost) is shown in the {\it left panel} and case II (with boost) is shown in the {\it right panel}. The labels denote the values of R.
 
Figure~\ref{fig:ratio} shows that the infalling gas erases part of the Ly$\alpha$ line redward of the Ly$\alpha$ resonance, and produces a sharp cut-off in the Ly$\alpha$ line at $x>-10$ \citep[see][for a more detailed discussion of this]{paperI}. For case I, $R=1.03$, which is outside the acceptable range found in \S~\ref{sec:UV}. For case II however, the boost in the ionising background reduces the neutral fraction in the IGM, which allows a small fraction of Ly$\alpha$ to 'leak' through the IGM. In this case the total fraction of Ly$\alpha$ that is transmitted depends quite strongly on the mean density of hydrogen, which results in $R=1.3$, well within the acceptable range for $R$ and less than $2\sigma$ away from the best-fit model if the UV-LF of LAEs were identical at $z=5.7$ and $z=6.5$.

\subsection{Constraints from $R\equiv \mathcal{T}_{\alpha,57}/\mathcal{T}_{\alpha,65}<1.8$}
\label{sec:bub}

\begin{figure}
\vbox{\centerline{\epsfig{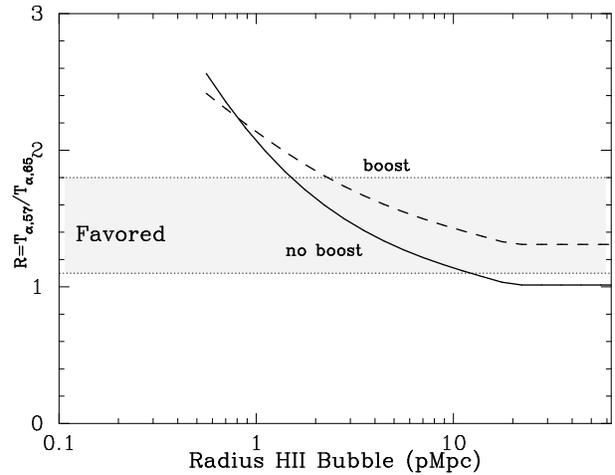}}}
\caption[]{The expected value of $R\equiv \mathcal{T}_{\alpha,57}/\mathcal{T}_{\alpha,65}$ for cases in which galaxies at $z=6.5$ are embedded within bubbles of reionised gas of radius $R_{\rm HII}$, surrounded by a fully neutral IGM. The {\it grey area} at $1.1 < R <1.8$ denotes the region of parameter-space that is favored by the joint UV and Ly$\alpha$ LFs. The {\it dashed (solid) line} show cases in which the ionising background is (is not) boosted by undetected surrounding sources. Existing data suggests that $R_{\rm HII}\gsim 2$ pMpc, which implies that the Universe is more than half ionised by volume at $z=6.5$ (see text).}
\label{fig:bub}
\end{figure}

In the previous section we found that the ratio of transmissions $R=\mathcal{T}_{\alpha,57}/\mathcal{T}_{\alpha,65}$ expected for a fully reionised IGM out to $z \geq 6.5$ can be as large as $R=1.3$. Following the analysis presented in \citet{paperI}, we calculate the expected value of $R$ for cases in which galaxies at $z=6.5$ are embedded in bubbles of reionised gas of radius $R_{\rm HII}$, within an otherwise neutral IGM. Prior to overlap, HII bubbles are expected to be generated by clusters of ionising sources. Their sizes therefore greatly exceed that of HII regions of individual galaxies \citep{F04a,F04b,WL04b,Zahn06}. The associated value of $R$ is shown in Figure~\ref{fig:bub} as a function of bubble radius for cases in which the local ionising background is boosted (not boosted) by nearby undetected sources as the {\it dashed (solid) line}.

Figure~\ref{fig:bub} shows that $R$ decreases with bubble size ($R_{\rm HII}$), as the damping wing absorption becomes less important. The minimum bubble radius shown corresponds to the case of an isolated galaxy in its own HII region. For large bubble sizes the ratio levels off for the boosted (not boosted) case at $R=1.3$ ($R\sim 1.03$). These are the values quoted in \S~\ref{sec:ratio} in reference to an ionised IGM at higher density. The {\it light grey area} shows the range $R\in[1.1-1.8]$ that was preferred by the combination of the UV and Ly$\alpha$ LF (\S~\ref{sec:UV}). According to Figure~\ref{fig:bub}, the constraint $R<1.8$ translates to a minimum bubble radius of $R_{\rm HII}\gsim 2$ pMpc, which translates to $R_{\rm HII}\gsim 15$ cMpc. In the model of \citet{F06} the characteristic HII-bubble size at a given epoch is related to the globally averaged ionised fraction of the universe by volume, $x_{\rm i,V}$ at that epoch. If we assume that the LAEs at $z=6.5$ reside in HII-bubbles that are of characteristic size at $z=6.5$, then the constraint $R_{\rm HII}\gsim 15$ cMpc translates to $x_{\rm i,V}\gsim 0.8$. However, the actual lower limit on $x_{\rm i,V}$ is weaker: if the Universe were truly more than $80\%$ ionised by volume at $z=6.5$, then the neutral IGM surrounding the HII bubble surrounding the LAEs would be filled with other ionised bubbles, which would reduce the damping wing optical depth of the IGM, which would enhance $\mathcal{T}_{\alpha,65}$ and thus reduce $R$. If the damping wing optical depth is reduced by a factor of $2$ (which is roughly the case when the universe is half ionised by volume), then we find our lower limit on bubble size to be $R_{\rm HII}\gsim 6-10$ cMpc, which translates to $x_{\rm i,V}\gsim 0.5-0.6$. Note that this lower limit on $x_{\rm i,V}$ would also be obtained directly from the constraint $R_{\rm HII}\gsim 15$ cMpc, when the impact of feedback from clustered sources is included, which modifies the relation between $R_{\rm HII}$ and $x_{\rm i,V}$ \citep{Kramer06}. In summary, the upper limit on $R$ suggests that a more reasonable lower limit on the ionised fraction of the Universe is $x_{\rm i,V}\gsim 0.5$. 

\section{Discussion \& Conclusions}
\label{sec:conclusion}

Recent observations have shown that the observed number counts of Ly$\alpha$ emitters evolve significantly between $z=5.7$ and $z=6.5$ \citep{Ka06}. It has been suggested that this evolution could be due to a significant change in the ionisation state of the IGM during this short time interval. In this paper we have investigated the constraints that may be placed on the evolution of the IGM transmission using the observed UV and Ly$\alpha$ LFs at $z=5.7$ and $z=6.5$. 

We used a simple prescription to relate the Ly$\alpha$ luminosity of a galaxy to the mass of its host dark matter $M_{\rm tot}$ and found that such a model can reproduce the data quite well. Using this model, we have shown that the observed Ly$\alpha$ LFs at $z=5.7$ and $z=6.5$ are best described by a model in which the IGM transmission evolves only weakly between these two redshifts. In fact, it is possible to attribute the observed evolution in the Ly$\alpha$ LF entirely to the evolution in the mass function of dark matter halos. The presence of cosmic variance in the observations strengthens this conclusion.

However, the observed rest-frame UV-LF of Ly$\alpha$ emitters appears not evolve between $z=5.7$ and $z=6.5$. Accounting for the co-evolution implies that the observed evolution of the Ly$\alpha$ LF may indeed be due to the evolution of the IGM transmission between $z=6.5$ and $z=5.7$. We find in this case that the ratio of transmissions is $1.1<R\equiv \mathcal{T}_{\alpha,57}/\mathcal{T}_{\alpha,65}<1.8$ ($\sim95\%$ confidence levels). This result is insensitive to the underlying model of the Ly$\alpha$ LF (as well as cosmic variance). However, we find that the ratio of transmissions expected for a fully reionised IGM out to $z \geq 6.5$ can be as large as $R=1.3$. Thus as with consideration of the Ly$\alpha$ LFs alone, the observed evolution is consistent with no change in the ionisation state of the IGM. The existing LFs therefore do not provide evidence for overlap between $z=5.7$ and $z=6.5$. Furthermore, the upper limit $R\equiv \mathcal{T}_{\alpha,57}/\mathcal{T}_{\alpha,65}<1.8$ implies that the Universe at $z=6.5$ was more than half ionised by volume, i.e. $x_{\rm i,V}>0.5$.

The present paper is based on observed LFs that were derived from spectroscopic observations of $17$ $z=6.5$ and $28$ $z=5.7$ Ly$\alpha$ emitters. A larger sample of Ly$\alpha$ emitters at both redshifts will allow more stringent constraints to be placed on simple models such as those presented here. In particular, the existing LF covers 1 dex in luminosity only. A larger range in luminosities will be useful for {\it (i)} breaking model degeneracies and determining the duty cycle and {\it (ii)} determining the evolution of the IGM transmission $\mathcal{T}_{\alpha}$ to greater accuracy, especially when Ly$\alpha$ LFs are used in combination with the UV-LF of Ly$\alpha$ emitters. This will provide constraints on the epoch of reionization which are independent of those derived from quasar absorption studies. An improved determination of the observed Ly$\alpha$ LF will also allow constraints to be placed on more sophisticated models, that account for the impact of the interstellar medium on the Ly$\alpha$ emission from galaxies \citep[as in][]{HS99}, include galaxy clustering and account for scatter in quantities such as $f_*\mathcal{T}_{\alpha}$, $\epsilon_{\rm DC}$ etc. These issues will be addressed in future work.

{\bf Acknowledgments} We thank N. Kashikawa for his helpful comments regarding the UV-Luminosity function. MD thanks D. Stark for correspondence that led to fixing a small error in Eq~2. We thank the anonymous referee for useful comments, and Avi Loeb for useful discussions. MD \& JSBW thank the Australian Research Counsel for support and the Center for Astrophysics and the Astronomy Department at Columbia University for their hospitality. ZH acknowledges partial support by NASA through grant NNG04GI88G, by the NSF through grant AST 0307291, and by the Hungarian Ministry of Education through a Gy\"orgy B\'ek\'esy Fellowship.

\label{lastpage}
\end{document}